\documentclass[a4wide]{article}

\usepackage{graphicx}
\usepackage{amssymb} 
\usepackage{amsmath} 
\usepackage{amsthm}  
\usepackage{array} 
\usepackage[usenames,dvipsnames]{color} 
\usepackage{bbm} 
\usepackage{url} 
\usepackage{a4wide} 
\usepackage[citebordercolor={0.1 0.32 0.76}, linkbordercolor={0.1 0.7 0.3}]{hyperref} 
\usepackage{cite}

\usepackage{tikz}
\usetikzlibrary{arrows,calc}
\usepackage{relsize}
\usetikzlibrary{decorations.pathmorphing} 

\def\@xfootnote[#1]{%
  \protected@xdef\@thefnmark{#1}%
  \@footnotemark\@footnotetext}

\title{ 
The zeroth law of thermodynamics is redundant}
\author{\normalsize Philipp Kammerlander, Renato Renner \\
\normalsize{Institute for Theoretical Physics, ETH Z\"urich, 
Switzerland}}
\date{\normalsize \today}

\begin{document}

\maketitle

\begin{abstract}\noindent
We reconsider the basic building blocks of classical phenomenological thermodynamics.
While doing so we show that the zeroth law is a redundant postulate for the theory by deriving it from the first and the second laws.
This is in stark contrast to the prevalent conception that the three laws, the zeroth, first and second, are all necessary and independent axioms.

\end{abstract}

\subsection*{Introduction}

The beginnings of the theory of thermodynamics are more than 200 years old \cite{Clausius50, Rankine50, Kelvin51, Maxwell71}.
While the initial evolve\-ment of the theory advanced alongside the development of steam engines that where later used to power locomotives and other machines ubiquitous in the industry, thermo\-dynamics has since been applied to more and more complex systems over almost all orders of magnitudes in size and levels of complexity.
Countless books have been written and many more lectures have been given on the subject \cite{Fowler39, Planck14, Buchdahl66, Carnot24, Planck97, Caratheodory09, Feynman63, Giles64, Fermi56, Zemansky68, Pauli73, Adkins83, Neumaier07, Thess11, Hulse18}.

The two most common ways to introduce thermodynamics are the \emph{phenomenological} approach, in which one takes the laws of thermodynamics as postulates,
and the \emph{statistical} approach, where the macroscopic thermodynamic properties of a system 
and hence the laws of thermodynamics
are derived by analysing its microscopic degrees of freedom using methods from statistical physics, e.g.\ \cite{Salem06}.
This paper is concerned with the phenomenological paradigm only. Arguments involving microscopic properties of a system shall only be used to give intuition in certain examples, if at all.

In numerous discussions with colleagues the authors made the experience that their queasy feeling about the foundations of phenomenological thermodynamics, and in particular about how they are taught in undergraduate courses and in recognised books, is shared by many others. 
Interestingly enough, not much literature on systematic modern approaches to thermodynamics exists. An exception is \cite{LY99,LY02}.
Nevertheless, there are still many different views on the foundations of the theory and how it should be introduced. 
While not so important for classical applications of the theory, for instance in engineering, the different seemingly contradicting alternatives for its foundations are confusing and unsatisfactory from a conceptual point of view.

Another reason to look more deeply into the basics of phenomenological thermodynamics is the increasing interest in and the emergence of the new research field of quantum thermodynamics. 
Various ideas have been proposed in the recent past on how to describe small (quantum) systems thermodynamically. A particular focus has been the comparison of quantum with classical thermodynamics and the quest for finding thermodynamic applications where quantum systems can surpass the power of their classical counterparts. 
However, the community is still far from agreeing on the conclusions,
an example being the ongoing controversy about the definition of work in the presence of quantum coherent states \cite{Talkner07, Aberg14, Perarnau17}. We argue that this is not least so because a comparison of classical and quantum thermodynamics is bound by the limited understanding of the basics of the classical theory itself.\\

In this paper we sketch a new way of laying out the foundations of thermodynamics that starts from the very basic concepts, such as systems, processes and states. 
We find that the zeroth law as a postulate is not needed.
In contrast to the statistical approach, where one of the main goals is to derive the fundamental thermodynamic laws, this result may seem surprising in the phenomenological approach
as it is the common view that the zeroth, first and second laws are all necessary and independent axioms for the theory of thermodynamics.  \\

The zeroth law has a long history and can be phrased in many different ways \cite{Maxwell71, Planck14, Fowler39, Buchdahl66}. It states that the relation ``being in thermal equilibrium with'' is \emph{transitive}. This means that if two bodies\footnote{Planck used the term ``bodies'' in \cite{Planck14}. Other terms that have been used in this context are ``assemblies'' \cite{Fowler39} or ``systems'' \cite{Buchdahl66}.}
$A$ and $B$ are in thermal equilibrium with each other and likewise $B$ and $C$, then also $A$ and $C$ must be in thermal equilibrium with each other.
Together with the usually implicitly assumed reflexivity ($A$ is in thermal equilibrium with $A$) and symmetry (if $A$ is in thermal equilibrium with $B$, then also vice versa) it implies that this relation is an \emph{equivalence relation}. 

The fact that ``being in thermal equilibrium with'' is an equivalence relation
is important to introduce a meaningful notion of \emph{empirical} temperature. One can now say that two bodies that are in thermal equilibrium with each other have the same temperature. 

While the zeroth law is usually postulated along with the first and second law, we argue that the zeroth law is implied by the latter
albeit now for \emph{absolute} temperature rather than empirical temperature. 
Notice that it is not possible to abandon the first or second law of thermodynamics without making the theory irrecognisable. Hence the fact that postulating the zeroth law is not necessary implies a sharp distinction between the status of the zeroth versus the first and second laws of thermodynamics. \\

In order to present a convincing reasoning for the redundancy of the zeroth law it is necessary to state all assumptions explicitly to make sure that it is not introduced behind the scenes.
We start by specifying a basic set of principles and postulates that are needed to talk about thermodynamic systems and processes. 
As we shall see this set can be kept rather small. 
For instance, we do not require an \emph{a priori} notion of ``thermal equilibrium''. Taking the zeroth law as a postulate would hence not even be possible.

\begin{figure}
\hskip-1cm
	\begin{tikzpicture}[scale=0.75, every node/.style={transform shape}]	
	
	\draw[red, fill=red!30!white, opacity=0.5] 
	(-4.25,3) node[above, black, opacity=1] {def.\ empirical}
	node[below, black, opacity=1] {temperature} ellipse (1.5cm and .7cm);
	\draw[blue, fill=blue!20!white, opacity=0.5] 
	(-1.25,3) node[black, opacity=1] {first law} ellipse (1cm and .5cm);
	\draw[blue, fill=blue!20!white, opacity=0.5] 
	(1.25,3) node[black, opacity=1] {second law} ellipse (1cm and .5cm);
	
	\draw[->] (-4.7,1.7) -- (-4.5,2.2);
	\draw[->] (-3.4,2.4) -- (-2.5,1.4);
	\draw[->] (-1.25,2.5) -- (-1.25,1.55);
	\draw[->] (1,2.5) -- (0,1.4);

	\draw[blue, fill=blue!20!white, opacity=0.5] 
	(-5,1) node[above, black, opacity=1] {postulated} 
	node[below, black, opacity=1] {zeroth law} ellipse (1.5cm and .7cm);
		
	\draw[green!80!black, fill=green!80!black, fill opacity=0.3] 
	(-1.25,1) node[black, opacity=1] {Carnot's theorem} ellipse (1.5cm and .5cm);	

	\draw[red, fill=red!30!white, opacity=0.5] 
	(2.5,1) node[above, black, opacity=1] {def.} 
	node[below, black, opacity=1] {reservoirs} ellipse (1.5cm and .7cm);
	
	\draw[->] (1,1) -- (.35,1);
	
	\draw[->] (-1,.5) -- (-.3,-.2);
	
	\draw[red, fill=red!30!white, opacity=0.5] 
	(-4.25,-1) node[above, black, opacity=1] {a priori def.} 
	node[below, black, opacity=1] {thermal equilibrium} ellipse (2cm and .7cm);
	\draw[red, fill=red!30!white, opacity=0.5] 
	(0,-1) node[above, black, opacity=1] {def.\ absolute} 
	node[below, black, opacity=1] {temperature} ellipse (1.5cm and .7cm);
	
	
	\draw[->] (-4.5,-.3) -- (-4.6,.25);

	\begin{scope}[xshift = 12cm]
		\draw[dotted] 
		(-4.25,3) node[above, black, opacity=1] {}
		node[below, black, opacity=1] {} ellipse (1.5cm and .7cm);
		\draw[blue, fill=blue!20!white, opacity=0.5] 
		(-1.25,3) node[black, opacity=1] {first law} ellipse (1cm and .5cm);
		\draw[blue, fill=blue!20!white, opacity=0.5] 
		(1.25,3) node[black, opacity=1] {second law} ellipse (1cm and .5cm);
	
		\draw[->] (-1.25,2.5) -- (-1.25,1.55);
		\draw[->] (1,2.5) -- (0,1.4);

		\draw[green!80!black, fill=green!80!black, fill opacity=0.3] 
		(-5,1) node[above, black, opacity=1] {derived} 
		node[below, black, opacity=1] {zeroth law} ellipse (1.5cm and .7cm);
		
		\draw[green!80!black, fill=green!80!black, fill opacity=0.3] 
		(-1.25,1) node[black, opacity=1] {Carnot's theorem} ellipse (1.5cm and .5cm);	

		\draw[red, fill=red!30!white, opacity=0.5] 
		(2.5,1) node[above, black, opacity=1] {def.} 
		node[below, black, opacity=1] {reservoirs} ellipse (1.5cm and .7cm);
	
		\draw[->] (1,1) -- (.35,1);
	
		\draw[->] (-1,.5) -- (-.3,-.2);
	
		\draw[red, fill=red!30!white, opacity=0.5] 
		(-4.25,-1) node[above, black, opacity=1] {def.\ thermal} 
		node[below, black, opacity=1] {equilibrium} ellipse (2cm and .7cm);
		\draw[red, fill=red!30!white, opacity=0.5] 
		(0,-1) node[above, black, opacity=1] {def.\ absolute} 
		node[below, black, opacity=1] {temperature} ellipse (1.5cm and .7cm);
	
		\draw[<-] (-2.15,-1) -- (-1.5,-1);
	
		\draw[->] (-4.5,-.3) -- (-4.6,.25);
	\end{scope}

	\end{tikzpicture}
\caption{Left: Overview of the reasoning in a standard line of argument \cite{Fermi56, Zemansky68, Pauli73, Adkins83}. Eventually, Carnot's theorem establishes the relation between the previously introduced empirical and absolute temperature. The color of the boxes indicates whether they are \textcolor{red}{definitions}, \textcolor{blue}{assumptions and postulates}, or \textcolor{green}{derived implications} in each setting.
Right: Overview of the reasoning presented in this paper.}
\label{fig:overview} 
\end{figure}
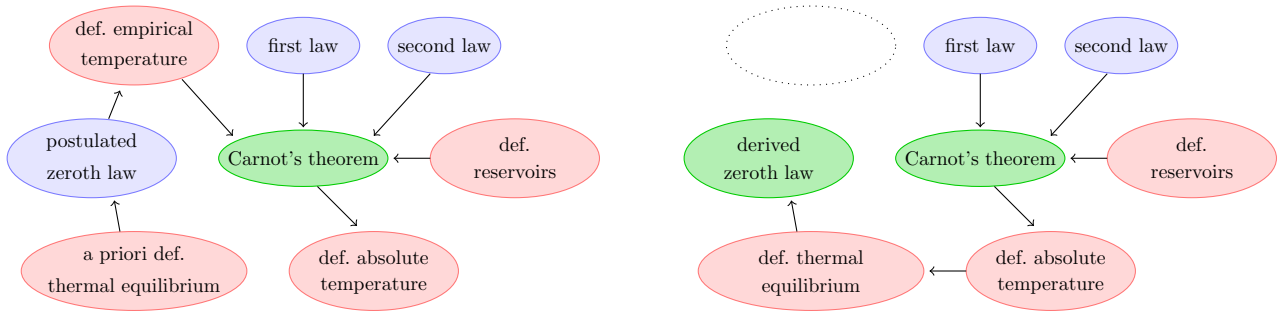

The essence of the argument lies in the proof of Carnot's theorem, similar to the standard proofs but without referring to the zeroth law and without making more assumptions than are usually made anyway, implicitly or explicitly. 
Using this result we introduce absolute temperature, a concept that can be used to define the notion of ``thermal equilibrium'', which we can then prove to satisfy the zeroth law.
Figure \ref{fig:overview} shows a schematic overview of the argument of this paper compared to standard approaches to the topic.






\ \\
\paragraph{Previous work.}
There exist other proposals on how to establish the theory of thermodynamics without the zeroth law, some of which are regrettably not very well known. 
Based on Carath\'eodory's version of the second law \cite{Caratheodory09} it was argued in \cite{Turner61, Turner63, Ehrlich81, Buchdahl86} that the zeroth law is not necessary as a postulate in thermodynamics. 
These results have been the topic of controversial discussions \cite{Miller52, Walter89, Buchdahl89, Turner05, Helsdon82, Clayton82}. 
The crucial difference to our approach is the order in which temperature and entropy is introduced. While in our approach entropy is a concept that follows after establishing absolute temperature, in Carath\'eodory-like approaches temperature is derived from entropy, namely it is its derivative with respect to internal energy. 
For a more detailed discussion of similarities and differences see Figure \ref{fig:LYoverview}, in which the results of the paper at hand are contrasted with \cite{LY99} that serves as the modern prototype of these works. 


A paper that deserves special mention is \cite{Home77}. An argument resembling ours very much is presented in a rather informal manner.
In contrast to the work presented here the assumptions are not spelled out explicitly and remain therefore somewhat unclear.
In addition, the concept of equilibrium is still used as an \emph{a priori} notion before absolute temperature is introduced. 
However, the sketched conclusions are based on similar ideas.

%


\subsection*{The argument}

Our goal is to give a proof of the main claim of this paper that 
is precise, but keep the technical formalism at a minimum.
We start with the basic concepts which, arguably, express standard assumptions 
that are usually made when thermodynamics is used to describe a system's physical behaviour. 
While they are stated in a rather abstract way in the text, Table \ref{tab:example} lists the concepts together with different standard examples for an easy intuitive understanding.

\paragraph{Basic concepts.} A thermodynamic theory is a theory making statements about a \emph{set of systems} $\mathcal{S}$. 
A \emph{thermodynamic process} $p$ can act on systems from this set. The set of thermodynamic processes is denoted by $\mathcal{P} $.
For any system $S\in\mathcal{S}$ the \emph{invested work} in a thermodynamic process $p$, $W_S(p)$, is described by a function 
$W_S: \mathcal{P} \rightarrow \mathbbm{R}$.
The function $W_S$ is additive under \emph{concatenation} of processes, meaning that when two processes are performed one after the other the total work cost is the sum of the work costs of the individual steps.
How to compute the work $W_S(p)$ of a certain process $p$ does not follow from the thermodynamic theory. On the contrary, it is something that the theory takes as an input. 
Typically the work is computed using a more fundamental theory such as classical mechanics, electrodynamics or maybe even quantum mechanics. 
For examples see Table \ref{tab:example}.

If a process $p$ acts on a given system $S\in\mathcal{S}$, but does not involve any other systems, then we say that it is a \emph{work process for $S$}. For any system $S\in\mathcal{S}$ the work processes for $S$ are summarized in the set $\mathcal{P}_S$. 
In a work processes for $S$ work can be invested or drawn from the system.
These processes are sometimes referred to as \emph{adiabatic} \cite{LY99}, however, here we call them work processes in order not to confuse terms.\footnote{We do not want to invoke the term \emph{adiabatic} at this point because it is usually used to denote processes that do not lead to heat flows. Since the concept of heat has not yet been introduced it is too early to define it as such.} 

Finally a process also specifies the \emph{initial and final states} of any system involved. This is again characterized by functions $\sigma_S^{\rm in/out}: \mathcal{P} \rightarrow \Sigma_S$ that exist for all systems in the set $\mathcal{S}$. 
The state space of a system $S$ is the union of all possible outputs of $\sigma^{\rm in}_S$ and $\sigma^{\rm out}_S$ and is called $\Sigma_S$.

When a system $S$ undergoes a process $p\in\mathcal{P}$ we denote the change in any state dependent quantity $X_S$ (state function) by $\Delta X_S := X_S(\sigma_S^{\rm out}(p)) - X_S(\sigma_S^{\rm in}(p))$.
A process $p\in\mathcal{P}$ is called \emph{cyclic on} $S$ if $\sigma_S^{\rm in}(p) = \sigma_S^{\rm out}(p)$.
For any system $S$ a work process $p\in\mathcal{P}_S$ is called \emph{reversible} if there is another work process on $S$ that has swapped input and output states, i.e.\ if there exists $p^{-1}\in\mathcal{P}_S$ with $\sigma_S^{\rm in}(p) = \sigma_S^{\rm out}(p^{-1})$ and $\sigma_S^{\rm in}(p^{-1}) = \sigma_S^{\rm out}(p)$.


\newpage

\vspace*{-3.2cm}
\begin{table}[h!]
\hskip-2cm
\begin{tabular}{p{2.3cm} | p{5cm} p{5.5cm} p{5cm}}
& $1$ mole of ideal gas, $S$ 
& two ideal gases, $S^{(2)} = c(S,S)$ 
& water tank, $N\gg1$ moles, $R'$ \\
\hline
intuitive \newline description 
& A container filled with an ideal gas. It is possible to read off the pressure inside the container and one can vary the volume by moving a piston.
& Two such ideal gases together. They can still be addressed individually but now they could also be thermally connected. The system is described by the composition $c(S,S)$.
& A very large water tank may be microscopically complex but is here described in the simplest non-trivial way, by its energy. It is an approximate reservoir. \\ 
 & & & \\
 state space  \newline $\Sigma_S$
& $\Sigma_{S} = (\mathbbm{R}^+)^2 \ni \sigma = (p,V)  $ \newline pressure and volume\protect\footnotemark 
& $\Sigma_{S^{(2)}} = (\mathbbm{R}^+)^4 \ni \sigma = (p_1,V_1,p_2,V_2)$ \newline both pressures and volumes 
& $\Sigma_{R'} = [E_{\rm min},E_{\rm max}] \ni \sigma = E$ \newline (internal) energy\protect\footnotemark \\
 & & & \\
processes \newline (examples \newline from $\mathcal{P}$) 
& $\bullet$ connecting the gas thermally \newline
\phantom{$\bullet$} to another system (irreversible \newline
\phantom{$\bullet$} in general) \newline 
\phantom{$\bullet$} $(p_{\rm in}, V_{\rm in}) \mapsto (p_{\rm out}, V_{\rm out}=V_{\rm in})$ \newline 
$\bullet$ isothermal expansion or com- \newline
\phantom{$\bullet$} pression (reversible, if done \newline
\phantom{$\bullet$} optimally) \newline 
\phantom{$\bullet$} $(p_{\rm in}, V_{\rm in}) \mapsto (p_{\rm out}, V_{\rm out})$ s.t.\newline \phantom{$\bullet$} \,$p_{\rm in}V_{\rm in} = p_{\rm out}V_{\rm out}$  \newline
$\bullet$ letting the gas undergo a \newline
\phantom{$\bullet$} Carnot cycle
& $\bullet$ all processes from the left column \newline
\phantom{$\bullet$} applied to one of the gases individ- \newline
\phantom{$\bullet$} ually \newline 
$\bullet$ thermally connecting the first gas \newline
\phantom{$\bullet$} to the second and the second to a \newline
\phantom{$\bullet$} reservoir (irreversible) 
& 
$\bullet$ connecting the tank thermally \newline 
\phantom{$\bullet$} to another system, thereby let- \newline 
\phantom{$\bullet$} ting them exchange energy (ir- \newline 
\phantom{$\bullet$} reversible in general) \newline
$\bullet$ using the tank in a thermal en- \newline 
\phantom{$\bullet$} gine together with a cyclic ma- \newline
\phantom{$\bullet$} chine and another tank \\
 & & & \\
work processes \newline (examples \newline from $\mathcal{P}_S$) 
& $\bullet$ shaking the gas, or applying \newline
\phantom{$\bullet$} friction to it (irreversible) \newline 
\phantom{$\bullet$} $(p_{\rm in}, V_{\rm in}) \mapsto (p_{\rm out}, V_{\rm out}=V_{\rm in})$ \newline 
\phantom{$\bullet$} with $p_{\rm out} > p_{\rm in}$ \newline 
$\bullet$ expanding or compressing \newline
\phantom{$\bullet$} the otherwise perfectly iso- \newline
\phantom{$\bullet$} lated gas\protect\footnotemark (reversible, if done \newline
\phantom{$\bullet$} optimally) \newline 
\phantom{$\bullet$} $(p_{\rm in}, V_{\rm in}) \mapsto (p_{\rm out}, V_{\rm out})$ s.t. \newline \phantom{$\bullet$} $p_{\rm in}V_{\rm in}^\gamma = p_{\rm out}V_{\rm out}^\gamma$ 
& $\bullet$ all processes from the left column \newline
$\bullet$ thermally connecting the two \newline
\phantom{$\bullet$} gases (irreversible in general) \newline
$\bullet$ isolating the gases and compress- \newline
\phantom{$\bullet$} ing one while expanding the other \newline
\phantom{$\bullet$} (reversible) \newline
$\bullet$ applying friction to one of the \newline
\phantom{$\bullet$} gases while thermally connecting \newline
\phantom{$\bullet$} them, thereby heating up the \newline
\phantom{$\bullet$} other one, too (irreversible) 
& $\bullet$ raising the internal energy of \newline
\phantom{$\bullet$} $R'$  by investing work: 
this \newline
\phantom{$\bullet$} could be done for instance by \newline
\phantom{$\bullet$} shaking it, applying friction, \newline
\phantom{$\bullet$} investing electrical work by let- \newline 
\phantom{$\bullet$} ting a current flow through a \newline 
\phantom{$\bullet$} resistance\\
 & & & \\
work function & For classical mechanical systems the infinitesimal work invested is always $\delta W = \vec{F}\cdot{\rm d}\vec{s}$. When shaking the gas or applying friction $\vec{F}$ is the applied force and d$\vec{s}$ the line segment of the trajectory along which the force acts. If the volume is changed without friction the infinitesimal work done can be simplified to $\delta W = -p{\rm d}V$.
& The total invested work of the composite system is the sum of the invested amounts of work on each individual system, $\delta W = \delta W_1 + \delta W_2$, where $\delta W_i$ is defined as in the left column for both gases individually.
& Depending on the  mechanism that is used to heat up the tank the work function will look different. For instance, in the case of increasing the internal energy by means of an electrical current $I$ flowing through a resistance $Z$ the invested work is $W = \int ZI^2 {\rm d}t$.

\end{tabular}
\caption{The basic concepts such as states, state spaces, processes, work processes, composition of systems, work cost functions and (an approximation of) reservoirs are illustrated with simple examples.}
\label{tab:example}
\end{table}

\addtocounter{footnote}{-2}
\footnotetext{We think of a truly ideal gas that remains an ideal gas even if $V$ becomes very small and $p$ very high. In the case of a real gas one could restrict the state space further, e.g.\ to $\Sigma_S = (0,p_{\rm max})\times(V_{\rm min},\infty)$.}


\addtocounter{footnote}{1}
\footnotetext{If one wants to use the tank as a reservoir, think of the energy range $[E_{\rm min},E_{\rm max}]$ as a small interval in a much larger spectrum of the system. As long as the energy is in this interval, the tank can exchange energy with other systems while keeping its properties and behaviour relative to other systems invariant (to good approximation). Intuitively one can always achieve this by taking a large enough system. 
When taking the limit towards an infinitely large water tank, $N\rightarrow\infty$ such that $\tfrac{N}{V} = \rm const.$ ($V$ the volume) $R'$ becomes an exact reservoir. 
For instance, when choosing $E_{\rm min / max} \propto \pm\sqrt{N}$,
in this limit the state space $\Sigma_{R'}$ will be the real line, i.e.\ $E_{\rm min / max} \rightarrow \mp\infty$ but still occupy only a very small part of the actual spectrum of the tank.
Reservoirs should be thought of as infinitely large systems that do not change their behaviour when exchanging finite amounts of energy with other systems.}

\addtocounter{footnote}{1}
\footnotetext{A compression or expansion of a perfectly isolated ideal gas leads to the known state changes as given in the table, where $\gamma$ is the heat capacity ratio $\tfrac{c_p}{c_V}$.}
\newpage

\paragraph{Composite systems.} When two separate systems are described as one, a new system is obtained. We say that any two systems $S_1\in\mathcal{S}$ and $S_2\in\mathcal{S}$ \emph{compose} a new system $c(S_1,S_2)\in\mathcal{S}$. In particular, we require that the set $\mathcal{S}$ is closed under composition. 
Moreover, the composition function $c$ is both associative and commutative. 
These properties then allow us to compose more than two systems by nesting such that e.g.\ for three systems $c(S_1,S_2,S_3) := c(c(S_1,S_2),S_3)$ is well-defined.

Composing two systems does not mean that they are thermally or otherwise coupled. They are simply described as one, akin to the tensor product structure from quantum theory, where $\mathcal{H}_1\otimes\mathcal{H}_2$ allows for a composite description of the two systems with associated Hilbert spaces $\mathcal{H}_1$ and $\mathcal{H}_2$.
For a process $p$ acting on both systems the work cost should again be additive in the sense that $W_{c(S_1,S_2)}(p) = W_{S_1}(p)+W_{S_2}(p)$. An illustrative example is given in the second column of Table \ref{tab:example}.

Any thermodynamic process on a system $S$ can be seen as a work process on a larger system, similar to the Stinespring dilation known from quantum information theory.
Notice that we do not need to define what a reversible thermodynamic process is. The definition of a reversible work process is sufficient. Since a thermodynamic process is a work process on a larger system, the reversibility of the former can be deduced from the reversibility of the latter.


\paragraph{First law.} The first law essentially guarantees that for any system $S$ there exists an \emph{internal energy} function $U_S:\Sigma_S\rightarrow\mathbbm{R}$. 
It does so by requiring two things: First, for any two states $\sigma_1,\sigma_2\in\Sigma_S$ there must be a work process $p\in\mathcal{P}_S$ that transforms $\sigma_1$ into  $\sigma_2$ or vice versa. Second, if two work processes on the same system transform the same input state into the same output state, then their work cost must be the same. 
Together these requirements make sure that an internal energy function exists for every system $S$ that satisfies $\Delta U_S = W_S(p)$ for any work process $p\in\mathcal{P}_S$.
From this together with additivity of $W_S$ under concatenation of processes it is easy to see that for a work process $q\in\mathcal{P}_S$ that leaves the initial state unchanged it holds $W_S(q)=0$. 
In a similar manner it follows that if $p^{-1}\in\mathcal{P}_S$ is the reverse process for $p\in\mathcal{P}_S$, then $W_S(p^{-1}) = -W_S(p)$. 

Obviously, the work cost on a system $S$ in a process that affects more systems than just $S$ does \emph{not only} depend on initial and final states. 
It depends on \emph{how exactly} the state change is brought about.\footnote{For example, heating up a box filled with gas may be done by applying friction, in which case the process has a non-zero work cost. 
However, the same state change from cold to warm while leaving the volume unchanged can be achieved by thermally connecting the gas to a warm reservoir. Certainly, this process can be carried out without investing nor drawing any work.} 
This means that in a process $p$ that is not a work process on $S$ the equation $\Delta U_S = W_S(p)$ will generally not be satisfied. We call the difference between the change in internal energy and the invested work \emph{heat}, $Q_S(p) := \Delta U_S - W_S(p)$. 
Seeing work as the energy we control, this definition allows us to interpret heat as the energy exchanges that we do not control -- energy that is exchanged between a heat bath and another system, for instance.

\paragraph{Heat reservoirs.} In order to formulate the second law according to Kelvin \cite{Kelvin51, Planck97} it is necessary to define the term \emph{heat reservoir} (also \emph{heat bath}) first. 
We characterize heat reservoirs by motivating intuitive assumptions that are 
formulated in terms of the previously introduces concepts.
First, a heat reservoir is usually thought of as a large but simple system. Large in the sense that it does not change its behaviour significantly if moderate amounts of energy are drawn from or given to it. Large also in the sense that it essentially does not matter whether one has access to one or two or more copies of the same reservoir. 
To this effect heat reservoirs are regarded as infinite systems.
Second, thermodynamically we describe heat reservoirs as simple systems in the sense that we do not want to consider more than one controllable macroscopic parameter with which one can change its state --
the only purpose is to provide or take up heat.\footnote{Since heat reservoirs are large systems they are of course not simple in a microscopic sense. On the contrary, we know that the larger the system, the more complex it gets when one tries to describe the interplay between its microscopic degrees of freedom. However, when saying simple, we here mean that the used \emph{thermodynamic description} is simple.} 

\begin{figure}
\begin{center} 
	\begin{tikzpicture}[scale=.65]
	\draw[] (-1,5.5) node[left] {(a)};
	
	\draw[very thick] (.5,4.5) node[above, yshift = .05cm] {$R$}-- (1,4.5) -- (1,5.5) -- (0,5.5) -- (0,4.5) -- (.5,4.5); 
	\draw[very thick] (.5,0) node[above, yshift = .05cm] {$R$} -- (1,0) -- (1,1) -- (0,1) -- (0,0) -- (.5,0); 
	\draw[] (.5,2.25) node[above, yshift = .05cm] {$S$} -- (1,2.25) -- (1,3.25) -- (0,3.25) -- (0,2.25) -- (.5,2.25);

	\draw[->] (.5,4.4) -- (.5,3.5) node [above right] {$Q_1$};
	\draw[->] (.5,1.1) -- (.5,2) node[below right] {$Q_2$};
	\draw[->] (2.15,2.75) -- (1.25,2.75) node[above right] {$W_S$};
	
	\begin{scope}[xshift = 8cm]	
	\draw[] (-3,5.5) node[left] {(b)};	
	
	\draw[very thick] (.5,5) node[] {$R$} circle (.6cm);
	\draw[very thick] (.5,.5) node[] {$R$} circle (.6cm);
	\draw[] (.5,2.25) node[above, yshift = .05cm] {$S$} -- (1,2.25) -- (1,3.25) -- (0,3.25) -- (0,2.25) -- (.5,2.25);

	\draw[->] (.5,4.3) -- (.5,3.5) node [above right] {$Q_1$};
	\draw[->] (.5,1.2) -- (.5,2) node[below right] {$Q_2$};
	\draw[->] (2.15,2.75) -- (1.25,2.75) node[above right] {$W_S$};
	
	\draw[very thick] (-2.5,2.25) node[above, yshift = .05cm] {$R$} -- (-2,2.25) -- (-2,3.25) -- (-3,3.25) -- (-3,2.25) -- (-2.5,2.25);
	\draw[->] (-1.9,2.75) -- (-1,5) node[left] {$Q_1$} -- (-.2,5);
	\draw[->] (-1.9,2.75) -- (-1,.5) node[left] {$Q_2$} -- (-.2,.5);
	
	\end{scope}
	
	\begin{scope}[xshift = 16cm]	
	\draw[] (-3,5.5) node[left] {(c)};	
	
	\draw[very thick] (.5,5) node[] {$R$} circle (.6cm);
	\draw[very thick] (.5,.5) node[] {$R$} circle (.6cm);
	\draw[] (.5,2.25) node[above, yshift = .05cm] {$S$} -- (1,2.25) -- (1,3.25) -- (0,3.25) -- (0,2.25) -- (.5,2.25);

	\draw[->] (2.15,2.75) -- (1.25,2.75) node[above right] {$W_S$};
	
	\draw[very thick, xshift=-.4cm] (-2.5,2.25) node[above, yshift = .05cm] {$R$} -- (-2,2.25) -- (-2,3.25) -- (-3,3.25) -- (-3,2.25) -- (-2.5,2.25);
	\draw[->] (-2.3,2.75) node[below right] {$Q_1+Q_2$} -- (-.2,2.75);
	
	\end{scope}
	
	\end{tikzpicture}
\end{center}
\caption{(a) An arbitrary system $S$ interacts with two copies of a reservoir $R$. (b) \& (c) Assumption (iii) for reservoirs states that the heat flows from the previous copies of the reservoirs can be supplied by another one of the same kind. The internal energy of the initial reservoirs has no net change, thus by (i) their states do not change either. The cyclicity is pictorially marked as circles around the reservoirs. Importantly, the two pictorial representations (b) and (c) of the internal heat flows are, according to our framework, thermodynamically equivalent (see the paragraph on the pictorial representation of heat flows).}
\label{fig:reservoir}
\end{figure}
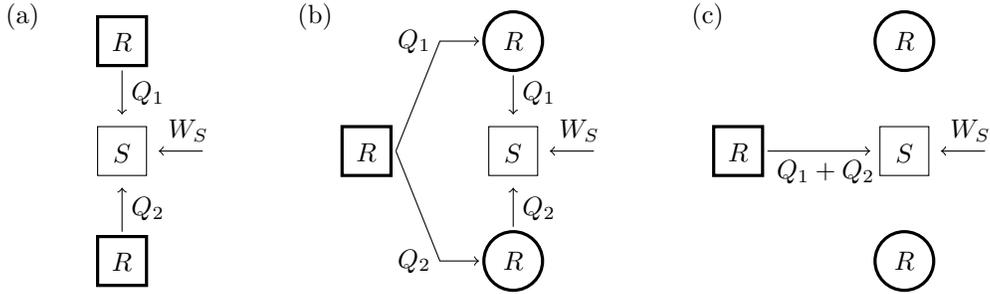

More formally, we use three postulates to characterize heat baths. For this, let $R\in\mathcal{S}$ be a system described by the thermodynamic theory. Then $R$ is called a heat reservoir if 
\begin{itemize}
	\item [(i)] 
	Its internal energy function $U_R$ is injective. 
	\item [(ii)] 
	There is no work process in $\mathcal{P}_R$ that extracts work from $R$.
	\item [(iii)] 
	If an arbitrary system $S\in\mathcal{S}$ interacts with two copies of the same heat reservoir, then the two reservoirs can be replaced by one single reservoir of the same kind that provides the sum of the heat flows that the previous copies supplied. See Figure \ref{fig:reservoir} for a
	pictorial illustration.
\end{itemize}

The first point states that the description of thermodynamic states of a heat reservoir is simple, namely that the states are in one-to-one correspondence with the internal energy 
and no other macroscopic quantity is needed to describe it. 
Point (ii) makes sure that a heat reservoir alone cannot produce work.
This is an important difference in comparison with work reservoirs. Work reservoirs are sometimes used when an explicit system is needed to model the exchange of work among systems. 
Finally, point (iii) states that a heat reservoir is a large systems in the sense motivated in Figure \ref{fig:reservoir}.

Arguably, the formulated assumptions on heat reservoirs are neither 
new nor surprising. They are mostly standard assumptions, see e.g.\ \cite{LY99}, that may not even be spelled out in certain texts on thermodynamics. 
However, here they are central for a precise argument. \\

One may wonder whether the zeroth law is already hidden in these assumptions together with the ones previously made. This is not the case. 
The one assumption most similar to the statement of the zeroth law is (iii). 
However, (iii) is strictly weaker than the zeroth law as stated in the introduction. The transitivity statement of the zeroth law is about three different systems and their relations whereas (iii) is about replacing two identical systems by another identical one. 
In fact, it is possible to construct a theory that satisfies all the assumptions made up until here  but nevertheless does not have a sensible zeroth law.
Only together with the last ingredient to come, the second law, will it be possible to formulate it. 



\paragraph{On the pictorial representation of heat flows.} When illustrating a process on a composite system we use thick lines for reservoirs and thin lines for other systems. 
Circles border cyclic systems while squares leave open whether the system involved undergoes cyclic evolution or not. 
Directed arrows mark positive work and heat flows. 
Arrows showing heat flows in more complex structures have no direct mathematical meaning since heat flows are not uniquely defined except when exchanged between two subsystems only. The internal arrows nevertheless help to map the abstract processes to well-known situations such as Carnot engines. 
For an example, see Figure \ref{fig:reservoir} (b) and (c). Both illustrations show the same process on the composite system $c(R,R,R,S)$. (b) suggests that the heat $Q_1$ and $Q_2$ flows from one copy of $R$ through the other ones to $S$ while (c) shows a direct heat flow of $Q_1+Q_2$ from one reservoir to $S$. 
The mathematical formalism leaves open which of the possibilities actually happen -- they are considered thermodynamically equivalent and both interpretations are valid. 


\paragraph{Second law.} Kelvin's second law states \cite{Kelvin51, Planck97}: ``It is impossible to devise a cyclically operating device, the sole effect of which is to absorb energy in the form of heat from a single thermal reservoir and to deliver an equivalent amount of work.'' 
With the notions and definitions made above we are now able to make this statement in a formal manner such that it can be used in the proof below. 

\begin{figure}[h]
\begin{center} 
	\begin{tikzpicture}[scale=.65]

	\draw[] (-1,5.5) node[left] {(a)};	

	\draw[very thick] (.5,4.5) node[above, yshift=.05cm] {$R$}-- (1,4.5) -- (1,5.5) -- (0,5.5) -- (0,4.5) -- (.5,4.5); 
	\draw[] (.5,2.75) node[] {$S$} circle (.6cm);

	\draw[<-] (.5,4.3) -- (.5,3.4) node [above right] {$Q_R(p)$};
	\draw[->] (2.15,2.75) node[right] {$W_S(p)$} -- (1.25,2.75);
	
	\begin{scope}[xshift = 8cm]
		\draw[] (-1,5.5) node[left] {(b)};	

		\draw[very thick] (.5,4.5) node[above] {$R_1$}-- (1,4.5) -- (1,5.5) -- (0,5.5) -- (0,4.5) -- (.5,4.5); 
		\draw[very thick] (.5,0) node[above] {$R_2$} -- (1,0) -- (1,1) -- (0,1) -- (0,0) -- (.5,0); 
		\draw[] (.5,2.75) node[] {$S$} circle (.6cm);

		\draw[->] (.5,4.4) -- (.5,3.5) node [above right] {$Q_1$};
		\draw[->] (.5,1.1) -- (.5,2) node[below right] {$Q_2$};
		\draw[->] (2.15,2.75) -- (1.25,2.75) node[above right] {$W_S$};
	\end{scope}
	
	\end{tikzpicture}
\end{center}
\caption{(a) The setting for the Kelvin statement of the second law. (b) Typical setting of a Carnot engine.}
\label{fig:2lawcarnot}
\end{figure}
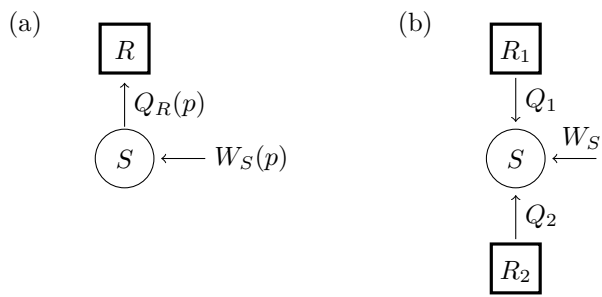

The setting is shown in Figure \ref{fig:2lawcarnot} (a). Let $R\in\mathcal{S}$ be a heat reservoir and $S\in\mathcal{S}$ an arbitrary system. 
Let furthermore $p\in\mathcal{P}_{c(S,R)}$ be a work process on the composite system $c(S,R)$ that satisfies $W_R(p)=0$ and $\sigma_S^{\rm in}(p) = \sigma_S^{\rm out}(p)$, i.e., no work is invested in nor drawn from $R$ and the process is cyclic on $S$. 
Then the second law states that $W_S(p) \geq0$. This is, no work can be drawn from such a cyclic machine interacting with a single heat reservoir.

 
Internal energy is a state function and due to the cyclicity of $S$ the internal energy $U_S$ does not change during the execution of $p$. Thus, the invested work is equal to the heat flowing from $S$ to $R$, $W_S(p) = Q_R(p)$. 
Hence the second law implies in the described setting that heat can only flow from $S$ to $R$, $Q_R(p) \geq0$.

\paragraph{Carnot's theorem.} The core of the argument presented in this paper lies in the proof of Carnot's theorem that is done without referring to the zeroth law. The setting relevant for Carnot's theorem is depicted in Figure \ref{fig:2lawcarnot} (b).
Let $S$ be a system operating cyclically between two heat reservoirs $R_1$ and $R_2$ (not necessarily of the same type). The system $S$ is from now on called machine. 
The three systems interact during a work process $p\in\mathcal{P}_{c(S,R_1,R_2)}$ such that $p$ is cyclic on $S$.
In addition, the setting shall be such that $W_{R_1}(p) = W_{R_2}(p) = 0$, i.e., the only non-zero work is invested into $S$ directly.
We call $Q_1 := -Q_{R_1}(p)$ the heat flowing from $R_1$ to $S$ and the same for $Q_2 := -Q_{R_2}(p)$ with respect to $R_2$, see Figure \ref{fig:2lawcarnot} (b).
Furthermore, we assume that not both of these heat flows are zero, otherwise the situation at hand is trivial. 

%
%
%
%

In Lemma 1 in the Appendix we prove that in this setting one of the heat flows $Q_1$ and $Q_2$ must be negative. Therfore, for reversible settings, i.e., settings in which $p$ is a reversible work process, one of the heat flows is negative and the other one is positive. 

Carnot's theorem then states the following: When comparing different machines operating between the same two reservoirs the ratio $-\tfrac{Q_1}{Q_2}$ is maximal for reversible processes $p$, where the numbering of the reservoirs is such that $Q_2$ is a negative heat flow. For any reversible work process $p$ the ratio is positive and depends only on the two reservoirs $R_1$ and $R_2$ and not on the 
machine nor the details of the process. 

The proof for Carnot's theorem is similar to standard proofs from textbooks of undergraduate courses, see e.g.\ \cite{Feynman63}. However, since we stated all necessary assumptions explicitly we are now in the position to claim that the proof does not make use of the zeroth law.

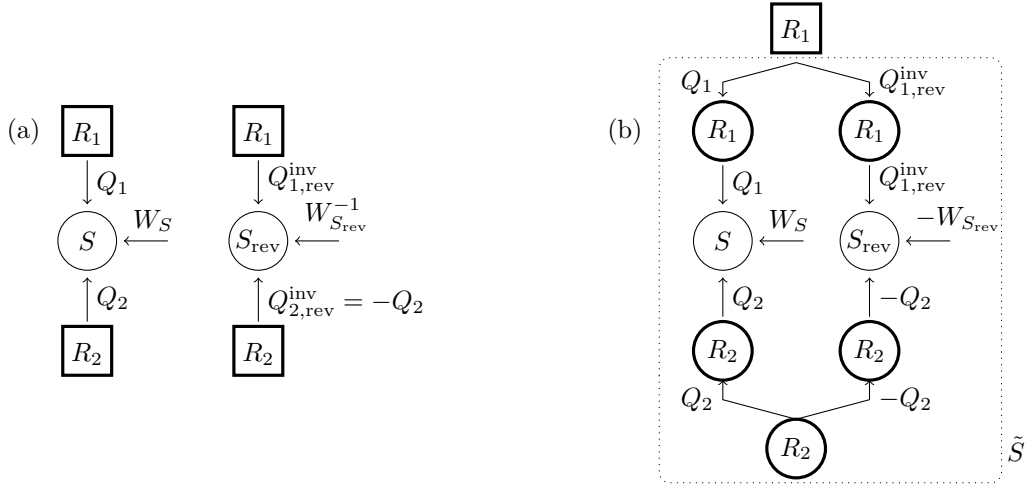
\begin{figure}[h]
\begin{center} 
	\begin{tikzpicture}[scale=.65]
	\draw[] (-.8,5) node[] {(a)};
	\draw[very thick] (.5,4.5) node[above] {$R_1$}-- (1,4.5) -- (1,5.5) -- (0,5.5) -- (0,4.5) -- (.5,4.5); 
	\draw[very thick] (.5,0) node[above] {$R_2$} -- (1,0) -- (1,1) -- (0,1) -- (0,0) -- (.5,0); 
	\draw[] (.5,2.75) node[] {$S$} circle (.6cm);

	\draw[->] (.5,4.4) -- (.5,3.5) node [above right] {$Q_1$};
	\draw[->] (.5,1.1) -- (.5,2) node[below right] {$Q_2$};
	\draw[->] (2.15,2.75) -- (1.25,2.75) node[above right] {$W_S$};
	
	\begin{scope}[xshift = 3.5cm]
		\draw[very thick] (.5,4.5) node[above] {$R_1$}-- (1,4.5) -- (1,5.5) -- (0,5.5) -- (0,4.5) -- (.5,4.5); 
		\draw[very thick] (.5,0) node[above] {$R_2$} -- (1,0) -- (1,1) -- (0,1) -- (0,0) -- (.5,0); 
		\draw[] (.5,2.75) node[] {$S_{\rm rev}$} circle (.6cm);
	
		\draw[->] (.5,4.4) -- (.5,3.5) node [above right] {$Q_{1,\rm rev}^{\rm inv}$};
		\draw[->] (.5,1.1) -- (.5,2) node[below right] {$Q_{2,\rm rev}^{\rm inv}=-Q_2$};
		\draw[->] (2.15,2.75) -- (1.25,2.75) node[above right] {$W_{S_{\rm rev}}^{-1}$};
	\end{scope}
	
	\begin{scope}[xshift = 13cm]
		\draw[] (-1.5,5) node[] {(b)};	
		\draw[very thick, xshift = 1.5cm, yshift = 2cm] (.5,4.6) node[above] {$R_1$}-- (1,4.6) -- (1,5.6) -- (0,5.6) -- (0,4.6) -- (.5,4.6); 
		\draw[->] (2,6.4) -- (.5,6) node[left] {$Q_1$} -- (.5,5.7);
		\draw[->] (2,6.4) -- (3.5,6) node[right] {$Q_{1,\rm rev}^{\rm inv}$} -- (3.5,5.7);
	
		\draw[very thick] (.5,5) node[] {$R_1$} circle (.6cm);
		\draw[very thick, xshift = 1.5cm, yshift = -2cm] (.5,.5) node[] {$R_2$} circle (.6cm);

		\draw[->] (2,-.9) -- (.5,-.5) node[left] {$Q_2$} -- (.5,-.1);
		\draw[->] (2,-.9) -- (3.5,-.5) node[right] {$-Q_2$} -- (3.5,-.1);
	
		\draw[very thick] (.5,.5) node[] {$R_2$} circle (.6cm);

		\draw[] (.5,2.75) node[] {$S$} circle (.6cm);

		\draw[->] (.5,4.3) -- (.5,3.5) node [above right] {$Q_1$};
		\draw[->] (.5,1.2) -- (.5,2) node[below right] {$Q_2$};
		\draw[->] (2.15,2.75) -- (1.25,2.75) node[above right] {$W_S$};
	
		\begin{scope}[xshift = 3cm]
			\draw[very thick] (.5,5) node[] {$R_1$} circle (.6cm);
			\draw[very thick] (.5,.5) node[] {$R_2$} circle (.6cm);
			\draw[] (.5,2.75) node[] {$S_{\rm rev}$} circle (.6cm);
	
			\draw[->] (.5,4.3) -- (.5,3.5) node [above right] {$Q_{1,\rm rev}^{\rm inv}$};
			\draw[->] (.5,1.2) -- (.5,2) node[below right] {$-Q_2$};
			\draw[->] (2.15,2.75) -- (1.25,2.75) node[above right] {$-W_{S_{\rm rev}}$};
	\end{scope}	\end{scope}
	
	\draw[dotted, xshift = 13cm, rounded corners] (-.8,0) -- (-.8,-2.2) -- (6.2,-2.2) -- (6.2,6.5) -- (-.8,6.5) -- (-.8,0);
	\draw[xshift = 13cm] (6.5,-1.5) node[] {$\tilde S$};

	\end{tikzpicture}
\end{center}
\caption{(a) Two Carnot engines, one of them reversible, run in parallel but are independent otherwise. (b) Using assumption (iii) on heat reservoirs twice, we can turn the situation from (a) to an equivalent situation in which there is one big cyclic machine $\tilde S$ interacting with a single heat reservoir $R_1$.}
\label{fig:carnot}
\end{figure}

\begin{proof}
The idea of the proof is to compare an arbitrary machine $S$ operating under an arbitrary work process $p\in\mathcal{P}_{c(S,R_1,R_2)}$ with a reversible one, $S_{\rm rev}$, operating under the reversible work process $p_{\rm rev}\in\mathcal{P}_{c(S_{\rm rev},R_1,R_2)}$. Using assumption (iii) for reservoirs we then couple two of the identical reservoirs to reduce the situation to the one from the second law. 

By definition $R_2$ is the reservoir that takes up heat from $S$ under $p$, i.e., $Q_2<0$, and $p_{\rm rev}$ is such that $Q_{2,\rm rev}:=Q_2(p_{\rm rev})<0$ as well such that the two machines work in the same ``direction''. This choice is up to us because $p_{\rm rev}$ is reversible by assumption. 
Furthermore, assume that $Q_{2,\rm rev} = Q_2$.\footnote{\label{ftnote:heat}If this was not the case one could take $k$ copies of machine $S$ and $l$ copies of machine $S_{\rm rev}$ such that $kQ_2 = lQ_{2,\rm rev}$ holds to arbitrary precision. In the end only the ratios of the heat flows matter and the number of copies of the machine cancel out, $-\tfrac{kQ_1}{kQ_2}=-\tfrac{Q_1}{Q_2}$.}
We will compare the ratios $-\tfrac{Q_1}{Q_2}$ and $-\tfrac{Q_{1,\rm rev}}{Q_{2,\rm rev}}$. In the proof we work with the inverse process on the reversible machine, $p_{\rm rev}^{-1}$, under which all signs of the heat and work flows change, $Q_{i,\rm rev}^{\rm inv} := Q_{i}(p_{\rm rev}^{-1}) = -Q_{i,\rm rev}$. Since only the signs change, the ratio $\tfrac{Q_1}{Q_2}$ stays the same.

When letting $p$ and $p_{\rm rev}^{-1}$ run in parallel we have a situation as described in Figure \ref{fig:carnot} (a) and can apply assumption (iii) for heat reservoirs twice, once for the two copies of $R_1$ and once for the two copies of $R_2$. 
By doing so we constructed a new situation as depicted in Figure \ref{fig:carnot} (b) in which one copy of $R_1$ interacts with a cyclic system $\tilde S = c(R_2,S,R_1,R_2,S_{\rm rev},R_1,R_2)$. This situation fits the one described in the second law and it follows that $Q_1+Q_{1,\rm rev}^{\rm inv} \leq0$. 
Rewriting this as $Q_1\leq -Q_{1,\rm rev}^{\rm inv}$ and multiplying it by $\tfrac{1}{-Q_2}>0$ we obtain
\begin{align}
-\frac{Q_1}{Q_2} \leq \frac{Q_{1,\rm rev}^{\rm inv}}{Q_2} = -\frac{Q_{1,\rm rev}^{\rm inv}}{Q_{2,\rm rev}^{\rm inv}} = -\frac{Q_{1,\rm rev}}{Q_{2,\rm rev}}\,.
\end{align}

If both machines $S$ and $S_{\rm rev}$ were reversible we could run the same argument with the corresponding reverse processes and obtain the inequality in the opposite direction. Therefore, in such a situation, the ratios must be equal. On top of that, for reversible machines the ratios are always positive because one of the heat flows is positive and the other one is negative. This concludes the proof. 
\end{proof}

\paragraph{Absolute temperature and the zeroth law.} Carnot's statement about the universality of the ratio of reversible heat flows allows us to introduce the notion of absolute temperature for heat reservoirs. 
For two heat reservoirs $R_1,R_2$ define $\tau(R_1,R_2):= -\tfrac{Q_1}{Q_2}$ to be the ratio for a reversible process, as discussed before, where again $Q_2<0$ is the negative heat flow.\footnote{By defining $\tau$ we make the implicit assumption that for any two reservoirs at least one machine exists that allows for a non-trivial reversible process as described before. Albeit not trivial this is a standard assumption when discussing the implications of Carnot's theorem.} 
Carnot's theorem guarantees that $\tau$ is well defined by certifying that the ratio does not depend on the machine nor the reversible process. 
In addition it says that for any two heat reservoirs $\tau(R_1,R_2)>0$. 

We briefly discuss different regimes in which reversible machines can operate.
If $\tau(R_1,R_2)>1$ the machine operates as a heat engine that produces work. In this case, the positive heat flow going into the machine is larger than the absolute value of the negative heat flow leaving the machine and the difference must be extracted as work. 
If $\tau(R_1,R_2)<1$ the opposite is the case: the machine works as a heat pump that consumes work. For $\tau(R_1,R_2)=1$ the machine simply transfers heat from one to the other heat reservoir and work is neither gained nor used.

In Lemma 2 in the Appendix we prove that for three heat reservoirs $R_1,R_2,R_3$ it always holds that
\begin{itemize}
	\item [(i)] 
	$\tau(R_1,R_1) = 1$,
	\item [(ii)]
	$\tau(R_2,R_1) = \tau(R_1,R_2)^{-1}$,
	\item [(iii)] 
	$\tau(R_1,R_2) \cdot \tau(R_2,R_3) = \tau(R_1,R_3)$.
\end{itemize}
We here give short intuitive arguments why these relations must hold. (i) When a reversible machine operates between two copies of the same reservoirs it can at most transfer heat from one to the other but it cannot produce work. This is a consequence of the second law. Therefore, the amount of heat flowing into one of the copies must be equal to the amount of heat flowing out of the other. (ii) When exchanging the roles of two reservoirs in the Carnot setting the two heat flows change the signs only. In the definition of $\tau$ we said that the negative heat flow must be in the denominator of the ratio, hence the ratio becomes its inverse. (iii) When considering two Carnot engines that share one heat reservoir ($R_2$) one can think of a new cyclic system consisting of the two Carnot engines and the shared heat reservoir. In essence this is a new Carnot setting with the remaining heat reservoirs ($R_1$ and $R_3$). Since the ratios of the heat flows are multiplicative, (iii) follows. \\

(i)--(iii) establish a relation on the set of heat reservoirs by means of $R_1\sim R_2 :\Leftrightarrow \tau(R_1,R_2)=1$. 
One could call this relation ``being in thermal equilibrium with'', which finally brings us to the zeroth law. Remember that the zeroth law states that such a relation must be an equivalence relation for it to make sense. In fact, the relation at hand is an equivalence relation: it is reflexive due to (i), symmetric due to (ii), and transitive due to (iii). 
However, we did not need to postulate this statement, on the contrary, we were able to derive it from existing postulates.

It is then possible to define absolute temperature by fixing a reference reservoir $R_{\rm ref}$ and assigning an arbitrary positive reference temperature $T_{\rm ref}$ to it. For any other reservoir $R$ its absolute temperature is defined as $T:= \tau(R,R_{\rm ref}) \cdot T_{\rm ref}$. 
For two arbitrary reservoirs $R_1,R_2$ with absolute temperature $T_1,T_2$ it follows $T_2 = \tau(R_2,R_1)\cdot T_1$, which shows that the choice of the reference reservoir is arbitrary. Furthermore, it implies the ultimate conclusion $R_1\sim R_2 \Leftrightarrow T_1=T_2$. \\

Absolute temperature introduced as such is only well-defined for heat reservoirs. At first sight this may seem a too narrow definition. 
However, according to the previous analysis we argue that this is the most one can hope for. We should not expect that temperature is a quantity that makes sense for an arbitrary system without further assumptions. 
Composite systems such as the one described in the second column of Table \ref{tab:example} are simple counter examples.
In a follow-up paper \cite{Kammerlander18} we show how a proper definition of the \emph{temperature of a heat flow} from or into 
an arbitrary system is nevertheless possible. 
In fact, this is the essential notion needed in order to state Clausius' theorem and define thermodynamic entropy along standard lines as the ratio between reversible heat and temperature \cite{Fermi56, Zemansky68, Pauli73, Adkins83}.


\subsection*{Discussion and outlook} 

We have shown that it is possible to define a notion of temperature, namely absolute temperature, without the necessity of postulating the zeroth law of thermodynamics. 
We did so by proving Carnot's theorem from basic standard assumptions, among them the first and second laws of thermodynamics. 
Using absolute temperature of reservoirs to define the equivalence relation ``being in thermal equilibrium with'' allows us to infer that the zeroth law holds nevertheless. Hence the zeroth law as a postulate is redundant. 

Nevertheless, the zeroth law seems a necessary postulate when the relation ``being in thermal equilibrium with'' is introduced before the second law. 
This is the case when one uses an \emph{a priori} notion of temperature, e.g.\ the (ideal) gas temperature discussed in \cite{Zemansky68}. 
But even if such an empirical temperature is introduced beforehand the theory of thermodynamics will eventually rely on the absolute temperature introduced via Carnot's theorem. 
In this sense the main statement of this paper is that two notions of temperature, empirical and absolute, is one too many -- the empirical temperature is unnecessary. If one waits with introducing temperature until the definition of absolute temperature the zeroth law will not be needed.  

Interestingly enough we did not have to make use of a definition of the term \emph{thermal equilibrium state} either. The stated assumptions seem to be sufficient to talk about thermodynamic states of systems without the use of this somewhat problematic \emph{a priori} definition. \\

\begin{figure}[h]
\hskip-1cm
	\begin{tikzpicture}[scale=0.75, every node/.style={transform shape}]	
	
	\draw[blue, fill=blue!20!white, opacity=0.5] 
	(-5.5,3) node[above, black, opacity=1] {postulates}
	node[below, black, opacity=1] {for $\succ$} ellipse (1.5cm and .7cm);
	\draw[blue, fill=blue!20!white, opacity=0.5] 
	(-2.5,3) node[black, opacity=1] {basic ass.} ellipse (1cm and .5cm);
	\draw[blue, fill=blue!20!white, opacity=0.5] 
	(0,3) node[black, opacity=1] {first law} ellipse (1cm and .5cm);
	\draw[green!80!black, fill=green!80!black, fill opacity=0.3] 
	(2.5,3) node[black, opacity=1] {second law} ellipse (1cm and .5cm);
	
	\draw[->] (-2.8,2.5) -- (-4.6,1.7);
	\draw[->] (-0.3,2.5) -- (-4,1.5);
	\draw[->] (-5.5,2.3) -- (-5.3,1.8);
	\draw[->] (-.5,1.5) -- (1.8,2.5);

	\draw[green!80!black, fill=green!80!black, fill opacity=0.3] 
	(-5.25,1) node[above, black, opacity=1] {entropy} 
	node[below, black, opacity=1] {principle} ellipse (1.5cm and .7cm);
		
	\draw[red, fill=red!30!white, opacity=0.5] 
	(-1.5,1) node[above, black, opacity=1] {def.\ entropy} 
	node[below, black, opacity=1] {as monotone} ellipse (1.5cm and .7cm);
	
	\draw[green!80!black, fill=green!80!black, fill opacity=0.3] 
	(2.25,1) node[above, black, opacity=1] {Carnot's} 
	node[below, black, opacity=1] {theorem} ellipse (1.5cm and .7cm);
	
	\draw[<-] (-3.1,1) -- (-3.75,1);
	
	\draw[->] (0,1) -- (.6,1);
		
	\draw[->] (-.8,.4) -- (0,-.4);
	
	\draw[dotted] 
	(-3,-1) node[above, black, opacity=1] {} 
	node[below, black, opacity=1] {} ellipse (1.5cm and .7cm);
	\draw[red, fill=red!30!white, opacity=0.5] 
	(1,-1) node[above, black, opacity=1] {def.\ absolute} 
	node[below, black, opacity=1] {temperature} ellipse (1.5cm and .7cm);
	
	\begin{scope}[xshift = 11.5cm]
		\draw[dotted] 
		(-5.5,3) node[above, black, opacity=1] {}
		node[below, black, opacity=1] {} ellipse (1.5cm and .7cm);
		\draw[blue, fill=blue!20!white, opacity=0.5] 
		(-2.5,3) node[black, opacity=1] {basic ass.} ellipse (1cm and .5cm);
		\draw[blue, fill=blue!20!white, opacity=0.5] 
		(0,3) node[black, opacity=1] {first law} ellipse (1cm and .5cm);
		\draw[blue, fill=blue!20!white, opacity=0.5] 
		(2.5,3) node[black, opacity=1] {second law} ellipse (1cm and .5cm);
	
		\draw[->] (-2.2,2.5) -- (1.1,1.6);
		\draw[->] (0.3,2.5) -- (1.7,1.8);
		\draw[->] (2.5,2.5) -- (2.4,1.8);

		\draw[green!80!black, fill=green!80!black, fill opacity=0.3] 
		(-5.25,1) node[above, black, opacity=1] {entropy} 
		node[below, black, opacity=1] {principle} ellipse (1.5cm and .7cm);
		
		\draw[red, fill=red!30!white, opacity=0.5] 
		(-1.5,1) node[above, black, opacity=1] {def.\ entropy} 
		node[below, black, opacity=1] {as $\Delta S = \tfrac{Q}{T}$} ellipse (1.5cm and .7cm);
	
		\draw[green!80!black, fill=green!80!black, fill opacity=0.3] 
		(2.25,1) node[above, black, opacity=1] {Carnot's} 
		node[below, black, opacity=1] {theorem} ellipse (1.5cm and .7cm);
	
		\draw[->] (-3,1) -- (-3.6,1);
	
		\draw[->] (2,.3) -- (1.7,-.3);
	
		\draw[red, fill=red!30!white, opacity=0.5] 
		(-3,-1) node[above, black, opacity=1] {def.\ temperature} 
		node[below, black, opacity=1] {of heat flows} ellipse (1.5cm and .7cm);
		\draw[red, fill=red!30!white, opacity=0.5] 
		(1,-1) node[above, black, opacity=1] {def.\ absolute} 
		node[below, black, opacity=1] {temperature} ellipse (1.5cm and .7cm);
	
		\draw[->] (-.5,-1) -- (-1.4,-1);
	
		\draw[->] (-3,-.3) -- (-2.5,.4);
	\end{scope}

	\end{tikzpicture}
\caption{The main line of argument as presented in \cite{LY99} (left) and in the framework used in this paper (right). The color code is the same as in Figure \ref{fig:overview}: \textcolor{red}{definitions}, \textcolor{blue}{assumptions and postulates}, and \textcolor{green}{derived implications}. Left: Defining an order relation $\succ$ and postulating several properties of it, Lieb and Yngvason are able to prove a theorem they call the ``entropy principle''. From there, the definition of entropy is obtained and it is possible to show that the second law holds as well as Carnot's theorem. 
Other approaches based on a second law similar to Carath\'eodoy's \cite{Caratheodory09} proceed in a similar way. 
Right: Overview of the reasoning presented in this paper and a follow-up paper \cite{Kammerlander18}. After introducing absolute temperature as presented before we will be able to define the absolute temperature of heat flows between arbitrary systems. Based on this, the usual definition of thermodynamic entropy can be used and theorems such as Clausius' theorem or the entropy principle can be proved. 
}
\label{fig:LYoverview}
\end{figure}
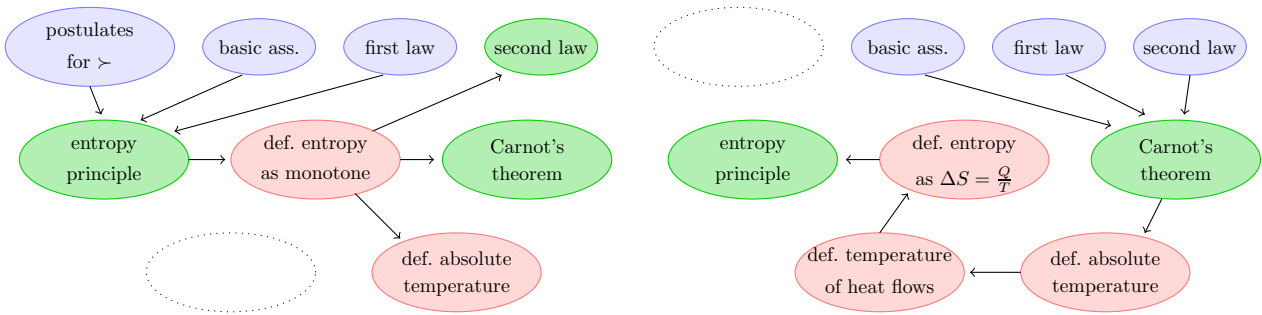

We conclude the paper with a brief outlook based on a comparison of our approach with the view proposed in \cite{LY99}. 
Eventually the definition of entropy is at the heart of a thermodynamic theory.
Figure \ref{fig:LYoverview} shows two diagrams with the main lines of argument presented in \cite{LY99} (left) and in this paper together with a follow-up paper \cite{Kammerlander18} (right).
One of the main differences lies in the definition of entropy. Whereas Lieb and Yngvason define it as a (unique) monotone, our approach sticks to the thermodynamic definition of entropy as the ratio of reversible heat divided by temperature. 
This is possible even though the term absolute temperature is not well-defined for arbitrary systems.
Lieb and Yngvason derive Carnot's theorem and different versions of the second law from their ``entropy principle'', which states the existence, monotonicity and additivity of the entropy function.
We go the other way around and first prove Carnot's theorem using the second law to define a thermodynamic entropy function that satisfies the entropy principle. 

The above observation reveals a further important difference between the two approaches. Lieb and Yngvason prove the second law starting from the postulates and assumptions on the order relation $\succ$, which implies that any thermodynamic theory described in their framework automatically satisfies the second law. 
The notion of thermodynamics without the second law is not even thinkable in this setting as it is part of the very basics of the theory. 
Therefore the statement ``One can derive the zeroth law from the first and second laws'', which is the main point of this paper, cannot even be made in the framework of Lieb and Yngvason. For such a statement both the first and second laws must enter the theory as an independent postulate. 
Hence the alternative approaches not only follow different paths in what is assumed and what is derived but also differ in the possible statements one can make.  

\paragraph{Acknowledgements.} P.K. thanks L\'idia del Rio and Tam\'as Kriv\'achy for helpful and controversial discussions that lead to significant improvements of this work. 
This work was funded by the Swiss National Science Foundation via project No.\ 200020\_165843 as well as the National Centre of Competence in Research ``Quantum Systems and Technology''.

\newpage
\subsection*{Appendix}

\paragraph{Lemma 1.}
Consider the setting described in Carnot's theorem, where a cyclic machine $S$ operates between two reservoirs $R_1$ and $R_2$. It is assumed that not both heat flows $Q_1$ and $Q_2$ are zero. The following statements hold:
\begin{itemize}
	\item [(i)] 	
	For arbitrary processes one of the heat flows must be negative.
	\item [(ii)]
	For reversible processes one of the heat flows is positive and the other one is negative.
\end{itemize}

The proof makes direct use of the requirements for heat reservoirs. This, together with another application in the proof of Carnot's theorem, is the place where the requirements are crucial. Notice that point (ii) proves that for reversible processes the ratio $-\tfrac{Q_1}{Q_2}$ must be positive. 

\begin{proof}\ \\
\vspace{-5mm}
\begin{itemize}
	\item [(i)]
	Let us first prove that if $Q_1=0$, then $Q_2<0$ (and likewise, if $Q_2=0$, then $Q_1<0$).
	Let $Q_1=0$. Since $\Delta U_{R_1}=Q_1=0$, the state of $R_1$ does not change. Therefore not only $S$ is cyclic but $\tilde S:= c(R_1,S)$ and the second law can be applied to $c(\tilde S,R_2)$. It follows that $0\geq Q_2$. Since we excluded the case that both heat flows are zero, we have $Q_2<0$. \\
	We now come to the main point of (i). Assume both heat flows are positive, i.e., heat flows from both reservoirs to the machine and is transformed into work. W.l.o.g.\ $Q_1\leq Q_2$. Due to the cyclicity of $S$ it holds $0=\Delta U = W_S + Q_1+Q_2$ and the invested work is $W_S = -Q_1-Q_2$. 
	We make use of requirement (ii) for heat reservoirs, which, together with the first law, implies that one can always invest positive work in a reservoir, thereby increasing its internal energy. Doing so with an amount of work equivalent to $Q_1$ with reservoir $R_1$ re-establishes its initial state due to requirement (i) for heat reservoirs. Therefore $c(S,R_1)$ is again a cyclic system and the second law can be applied. It implies that $Q_2\leq0$, which contradicts the assumption made in the beginning of the proof. \\
	\item [(ii)] Due to (i) one of the heat flows has to be negative. Suppose the other one was zero. If the reverse process exists it will lead to reversed heat flows, meaning that one heat flow will be positive and the other one zero. Such a situation still fits the setting we are considering here and thus contradicts (i). Therefore, for reversible machines, the heat that is not negative must be positive.
\end{itemize}
\end{proof}


\paragraph{Lemma 2.} 
Let $R_1,R_2,R_3$ be three heat reservoirs and $\tau$ the function defined in the main text. Then:
\begin{itemize}
	\item [(i)] 
	$\tau(R_1,R_1) = 1$,
	\item [(ii)]
	$\tau(R_2,R_1) = \tau(R_1,R_2)^{-1}$,
	\item [(iii)] 
	$\tau(R_1,R_2) \cdot \tau(R_2,R_3) = \tau(R_1,R_3)$.
\end{itemize}

\begin{proof}
We start by showing (iii), from which the other two statements follow.
\begin{itemize}
	\item [(iii)] 
	Let $S$ be a machine operating between $R_1$ and $R_2$ and $S'$ a machine operating between another copy of $R_2$ and $R_3$, see Figure \ref{fig:prooflemma2} on the left. Both machines shall operate via reversible processes. Further, let $Q_2<0$ be the negative heat flow of machine $S$ and w.l.o.g.\ $Q_2'=-Q_2>0$.\footnote{We do so here with the same argument as in the proof of Carnot's theorem, see Footnote \ref{ftnote:heat} on page \pageref{ftnote:heat}.}
	We then know that $Q_3'<0$ and thus, according to the definition of $\tau$, we find $\tau(R_1,R_2) = -\tfrac{Q_1}{Q_2}$ and $\tau(R_2,R_3) = -\tfrac{Q_2'}{Q_3'}$.
	
	As in the proof of Carnot's theorem we now use requirement (iii) for heat reservoirs and replace the two copies of $R_2$ with a third one that supplies the heat flows $Q_2$ and $Q_2'$. Since the internal energy changes in the now three copies of $R_2$ are all zero, their states are the same before and after the reversible processes. Hence they are also cyclic systems. 
	In total, we have a cyclic machine $\tilde S = c(R_2,S,R_2,S',R_2)$, as shown in Figure \ref{fig:prooflemma2} on the right, that operates between $R_1$ and $R_3$ by means of a work process. 
This process is constructed as a reversible work process. Its reverse is the same construction with the individual reverse work processes on the two machines $S$ and $S'$ together with the corresponding reservoirs. 
	By definition, we can now compute $\tau(R_1,R_3) = -\tfrac{Q_1}{Q_3'}$, which then satisfies $\tau(R_1,R_3) = \tau(R_1,R_2) \cdot \tau(R_2,R_3)$.

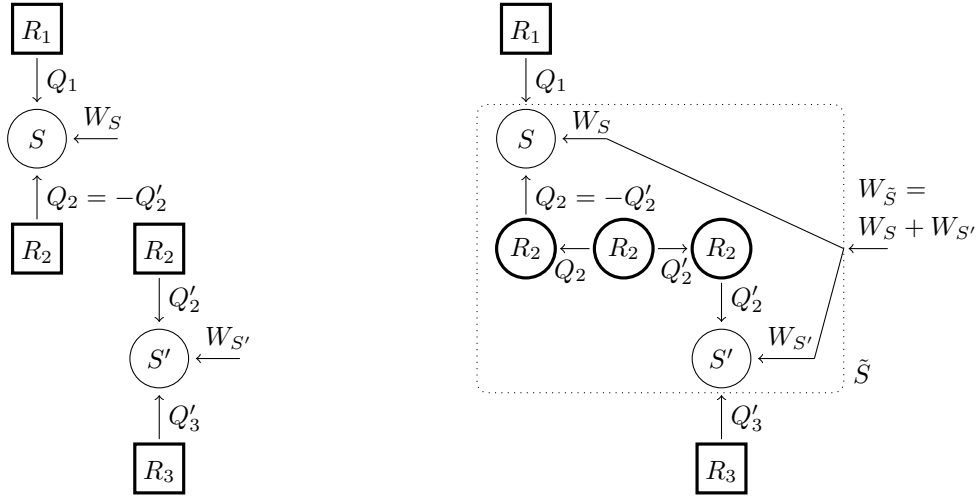
\begin{figure}
\begin{center} 
\begin{tikzpicture}[scale=.65]

	\draw[very thick] (.5,4.5) node[above] {$R_1$}-- (1,4.5) -- (1,5.5) -- (0,5.5) -- (0,4.5) -- (.5,4.5); 
	\draw[very thick] (.5,0) node[above] {$R_2$} -- (1,0) -- (1,1) -- (0,1) -- (0,0) -- (.5,0); 
	\draw[] (.5,2.75) node[] {$S$} circle (.6cm);

	\draw[->] (.5,4.4) -- (.5,3.5) node [above right] {$Q_1$};
	\draw[->] (.5,1.1) -- (.5,2) node[below right] {$Q_2 = -Q_2'$};
	\draw[->] (2.15,2.75) -- (1.25,2.75) node[above right] {$W_S$};
	
	\begin{scope}[xshift = 2.5cm, yshift = -4.5cm]
		\draw[very thick] (.5,4.5) node[above] {$R_2$}-- (1,4.5) -- (1,5.5) -- (0,5.5) -- (0,4.5) -- (.5,4.5); 
		\draw[very thick] (.5,0) node[above] {$R_3$} -- (1,0) -- (1,1) -- (0,1) -- (0,0) -- (.5,0); 
		\draw[] (.5,2.75) node[] {$S'$} circle (.6cm);
	
		\draw[->] (.5,4.4) -- (.5,3.5) node [above right] {$Q_2'$};
		\draw[->] (.5,1.1) -- (.5,2) node[below right] {$Q_3'$};
		\draw[->] (2.15,2.75) -- (1.25,2.75) node[above right] {$W_{S'}$};
	\end{scope}
	
	\begin{scope}[xshift = 10cm]
		
		\draw[very thick] (.5,4.5) node[above] {$R_1$}-- (1,4.5) -- (1,5.5) -- (0,5.5) -- (0,4.5) -- (.5,4.5); 
		\draw[very thick] (.5,.5) node[] {$R_2$} circle (.6cm);

		\draw[] (.5,2.75) node[] {$S$} circle (.6cm);

		\draw[->] (.5,4.4) -- (.5,3.5) node [above right] {$Q_1$};
		\draw[->] (.5,1.2) -- (.5,2) node[below right] {$Q_2 = -Q_2'$};
		\draw[->] (2.15,2.75) -- (1.25,2.75) node[above right, yshift=-.05cm] {$W_S$};
	
		\begin{scope}[xshift = 4cm, yshift = -4.5cm] 
			\draw[very thick] (.5,5) node[] {$R_2$} circle (.6cm);
			\draw[very thick] (.5,0) node[above] {$R_3$} -- (1,0) -- (1,1) -- (0,1) -- (0,0) -- (.5,0); 

			\draw[] (.5,2.75) node[] {$S'$} circle (.6cm);
	
			\draw[->] (.5,4.3) -- (.5,3.5) node [above right] {$Q_2'$};
			\draw[->] (.5,1.1) -- (.5,2) node[below right] {$Q_3'$};
			\draw[->] (2.4,2.75) -- (1.25,2.75) node[above right] {$W_{S'}$};
		\end{scope}
		\draw[very thick] (2.5,.5) node[] {$R_2$} circle (.6cm);
		\draw[->] (1.8,.5) node[xshift = -.25cm, yshift = -.3cm] {$Q_2$} -- (1.2,.5);
		\draw[->] (3.2,.5) node[xshift = .25cm, yshift = -.3cm] {$Q_2'$} -- (3.8,.5);
		
		\draw[dotted, rounded corners] (-.5,0) -- (-.5,2.1-4.55) -- (7,2.1-4.55) node[above right] {$\tilde S$} -- (7,3.45) -- (-.5,3.45) -- (-.5,0);
		
		\draw[] (2.4+4,2.75-4.5) -- (7,.5);
		\draw[] (2.15,2.75) -- (7,.5);
		\draw[->] (7.9,.5) -- (7.1,.5) node[above right] {$W_S + W_{S'}$};
		\draw[] (7.1,1.2) node[above right] {$W_{\tilde S} =$};
		
	\end{scope}

	\end{tikzpicture}
\end{center}	
\caption{Two individual reversible machines $S$ operating between the reservoirs $R_1$ and $R_2$ and $S'$ operating between $R_2$ and $R_3$ (left) are coupled to form one reversible machine $\tilde S$ operating between $R_1$ and $R_3$ (right).
}
\label{fig:prooflemma2}
\end{figure}
	
	\item [(i)] 
	Apply (iii) to $R_1=R_2=R_3$ and remember that $\tau$ is a positive-valued function.
	\item [(ii)]
	Choose $R_3=R_1$ and use again (iii). Together with (i) the claim follows.	
\end{itemize}
\end{proof}

\newpage
\addcontentsline{toc}{chapter}{References}
\bibliographystyle{unsrt}
\bibliography{references}

\begin{thebibliography}{10}

\bibitem{Clausius50}
R.~Clausius.
\newblock {\"U}ber die bewegende {K}raft der {W}\"arme und die {G}esetze,
  welche sich daraus fü\"ur die {W}\"armelehre selbst ableiten lassen.
\newblock {\em Annalen der Physik}, 79:368--397, 500--524, 1850.

\bibitem{Rankine50}
W.G. Rankine.
\newblock {\em {\"U}ber die mechanische {T}heorie der {W}\"arme}.
\newblock Wiley, 1850.

\bibitem{Kelvin51}
W.~Thomson.
\newblock On the {D}ynamical {T}heory of {H}eat, with numerical results deduced
  from {M}r. {J}oule's equivalent of a {T}hermal {U}nit, and {M}r. {R}egnault's
  {O}bservations on {S}team.
\newblock {\em Transactions of the Royal Society of Edinburgh}, XX:261--268,
  289--298, 1851.

\bibitem{Maxwell71}
J.C. Maxwell.
\newblock {\em The Theory of Heat}.
\newblock Longmans, Green, and Co., 1871.

\bibitem{Fowler39}
R.~Fowler and E.A. Guggenheim.
\newblock {\em Statistical Thermodynamics: A version of Statistical Mechanics
  for Students of Physics and Chemistry}.
\newblock Cambridge University Press, 1939.

\bibitem{Planck14}
M.~Planck.
\newblock {\em The Theory of Heat Radiation}.
\newblock P. Blakiston's Son \& Co., 1914.

\bibitem{Buchdahl66}
H.A. Buchdahl.
\newblock {\em The Concepts of Classical Thermodynamics}.
\newblock Cambridge University Press, 1966.

\bibitem{Carnot24}
S.~Carnot.
\newblock {\em R\'eflexions sur la puissance motrice du feu et sur les machines
  propres \`a d\'evelopper cette puissance}.
\newblock Bechelier, 1824.

\bibitem{Planck97}
M.~Planck.
\newblock {\em Vorlesungen \"uber Thermodynamik}.
\newblock Veit \& Comp., 1897.

\bibitem{Caratheodory09}
C.~Carath\'eodory.
\newblock {U}ntersuchung \"uber die {G}rundlagen der {T}hermodynamik.
\newblock {\em Math. Annalen}, 67:355--386, 1909.

\bibitem{Feynman63}
R.P. Feynman, B.~Leighton, and L.~Sands.
\newblock {\em The Feynman Lectures on Physics}.
\newblock Reading, Mass: Addison-Wesley Pub. Co, 1963.

\bibitem{Giles64}
R.~Giles.
\newblock {\em Mathematical Foundations of Thermodynamics}.
\newblock Pergamon, 1964.

\bibitem{Fermi56}
E.~Fermi.
\newblock {\em Thermodynamics}.
\newblock Dover, 1956.

\bibitem{Zemansky68}
M.W. Zemansky.
\newblock {\em Heat and Thermodynamics}.
\newblock McGraw-Hill Book Company, 1968.

\bibitem{Pauli73}
W.~Pauli.
\newblock {\em Thermodynamics and the {K}inetic {T}heory of {G}ases}.
\newblock MIT Press, Cambridge, 1973.

\bibitem{Adkins83}
C.J. Adkins.
\newblock {\em Equilibrium {T}hermodynamics}.
\newblock Cambridge University Press, 1983.

\bibitem{Neumaier07}
A.~Neumaier.
\newblock On the foundations of thermodynamics, 2007.
\newblock arxiv:0705.3790v1.

\bibitem{Thess11}
A.~Thess.
\newblock {\em The {E}ntropy {P}rinciple: {T}hermodynamics for the
  {U}nsatisfied}.
\newblock Springer-Verlag, Berlin, 2011.

\bibitem{Hulse18}
A.~Hulse, B.~Schumacher, and M.D. Westmoreland.
\newblock Axiomatic information thermodynamics, 2018.
\newblock arxiv:1801.05015.

\bibitem{Salem06}
W.K. Salem and J.~Fr\"ohlich.
\newblock Status of the {F}undamental {L}aws of {T}hermodynamics, 2006.
\newblock arxiv:math-ph/0604067v3.

\bibitem{LY99}
E.H. Lieb and J.~Yngvason.
\newblock The physics and mathematics of the second law of thermodynamics.
\newblock {\em Physics Reports}, pages 1--96, 1999.

\bibitem{LY02}
E.H. Lieb and J.~Yngvason.
\newblock The {M}athematical {S}tructure of the {S}econd {L}aw of
  {T}hermodynamics.
\newblock {\em Current Developments in Mathematics, 2001}, pages 89--130, 2002.

\bibitem{Talkner07}
Peter Talkner, Eric Lutz, and Peter H\"anggi.
\newblock Fluctuation theorems: {W}ork is not an observable.
\newblock {\em Phys. Rev. E}, 75:050102, 2007.

\bibitem{Aberg14}
Johan \AA{}berg.
\newblock Catalytic {C}oherence.
\newblock {\em Phys. Rev. Lett.}, 113:150402, 2014.

\bibitem{Perarnau17}
M.~Perarnau-Llobet, E.~B\"aumer, K.~Hovhannisyan, M.~Huber, and A.~Ac\'in.
\newblock No-{G}o {T}heorem for the {C}haracterization of {W}ork {F}luctuations
  in {C}oherent {Q}uantum {S}ystems.
\newblock {\em Phys. Rev. Lett.}, 118:070601, 2017.

\bibitem{Turner61}
L.A. Turner.
\newblock Zeroth {L}aw of {T}hermodynamics.
\newblock {\em American Journal of Physics}, 29:71, 1961.

\bibitem{Turner63}
L.A. Turner.
\newblock Temperature and {C}arath\'eodory's {T}reatment of {T}hermodynamics.
\newblock {\em The Journal of Chemical Physics}, 38:1163, 1963.

\bibitem{Ehrlich81}
P.~Ehrlich.
\newblock The concept of temperature and its dependence on the laws of
  thermodynamics.
\newblock {\em American Journal of Physics}, 49:622, 1981.

\bibitem{Buchdahl86}
H.A. Buchdahl.
\newblock On the redundancy of the zeroth law of thermodynamics.
\newblock {\em J. Phys. A: Math. Gen}, 19:L561--L564, 1986.

\bibitem{Miller52}
A.R. Miller.
\newblock The {C}oncept of {T}emperature.
\newblock {\em American Journal of Physics}, 20:488, 1952.

\bibitem{Walter89}
J.~Walter.
\newblock On {H} {B}uchdahl's project of a thermodynamics without empirical
  temperature as a primitive concept.
\newblock {\em J. Phys. A: Math. Gen.}, 22:341--342, 1989.

\bibitem{Buchdahl89}
H.A. Buchdahl.
\newblock Reply to comment by {J} {W}alter on `{O}n the redundancy of the
  zeroth law of thermodynamics'.
\newblock {\em J. Phys A: Math. Gen.}, 22:343, 1989.

\bibitem{Turner05}
L.A. Turner.
\newblock Further {R}emarks on the {Z}eroth {L}aw.
\newblock {\em American Journal of Physics}, 30:804, 2005.

\bibitem{Helsdon82}
R.M. Helsdon.
\newblock The zeroth law of thermodynamics.
\newblock {\em Phys. Educ.}, 17:114, 1982.

\bibitem{Clayton82}
D.G. Clayton and R.M. Helsdon.
\newblock Zeroth law of thermodynamics.
\newblock {\em Phys. Educ.}, 17:251, 1982.

\bibitem{Home77}
D.~Home.
\newblock Concept of temperature without the zeroth law.
\newblock {\em American Journal of Physics}, 45:1203, 1977.

\bibitem{Kammerlander18}
P.~Kammerlander and R.~Renner.
\newblock In preparation.

\end{thebibliography}

\end{document}